\documentclass[12pt, preprint]{aastex}
\usepackage{geometry} 
\usepackage{natbib}

\slugcomment{Accepted to ApJ}

\shorttitle{$H$-band Structure of P Cygni} 
\shortauthors{Richardson et al.} 

\begin{document}

\title{The $H$-band Emitting Region of the Luminous Blue Variable P Cygni: 
Spectrophotometry and Interferometry of the Wind}
\author{N. D. Richardson\altaffilmark{1,2,3}, 
G. H. Schaefer\altaffilmark{4},
D. R. Gies\altaffilmark{2,3},
O. Chesneau\altaffilmark{5},
J. D. Monnier\altaffilmark{6},
F. Baron\altaffilmark{6},
X. Che\altaffilmark{6},
J. R. Parks\altaffilmark{3},
R. A. Matson\altaffilmark{2,3},
Y. Touhami\altaffilmark{3},
D. P. Clemens\altaffilmark{7},\\
E. J. Aldoretta\altaffilmark{2,3},
N. D. Morrison\altaffilmark{8},
T. A. ten Brummelaar\altaffilmark{4}, 
H. A. McAlister\altaffilmark{3},\\
S. Kraus\altaffilmark{9},
S. T. Ridgway\altaffilmark{10},
J. Sturmann\altaffilmark{4},
L. Sturmann\altaffilmark{4},
B. Taylor\altaffilmark{7},\\
N. H. Turner\altaffilmark{4},
C. D. Farrington\altaffilmark{4}, and
P. J. Goldfinger\altaffilmark{4}}

\altaffiltext{1}{Current address: D\'epartement de physique and Centre de Recherche en Astrophysique du Qu\'ebec (CRAQ), Universit\'e de Montr\'eal, C.P. 6128, Succ.~Centre-Ville, Montr\'eal, Qu\'ebec, H3C 3J7, Canada; richardson@astro.umontreal.ca}
\altaffiltext{2}{Visiting Astronomer, Lowell Observatory}
\altaffiltext{3}{Center for High Angular Resolution Astronomy, Department of Physics and Astronomy, Georgia State University, P. O. Box 4106, Atlanta, GA  30302-4106, USA} 
\altaffiltext{4}{The CHARA Array, Georgia State University, P.O. Box 3965, Atlanta, GA 30302-3965, USA}
\altaffiltext{5}{Nice Sophia-Antipolis University, CNRS UMR 6525, Observatoire de la C$\hat{\rm o}$te d'Azur, BP 4229, 06304, Nice Cedex 4, France}
\altaffiltext{6}{Department of Astronomy, University of Michigan, 941 Dennison Bldg., Ann Arbor, MI 48109-1090, USA}
\altaffiltext{7}{Institute for Astrophysical Research, Boston University, 725 Commonwealth Ave, Boston, MA 02215, USA}
\altaffiltext{8}{Ritter Astrophysical Research Center, Department of Physics and Astronomy, University of Toledo, 2801 W. Bancroft, Toledo, OH 43606, USA}
\altaffiltext{9}{Harvard-Smithsonian Center for Astrophysics, 60 Garden Street, MS-78, Cambridge, MA 02138, USA}
\altaffiltext{10}{National Optical Astronomical Observatory, 950 North Cherry Ave., Tucson, AZ, 85719, USA}
\setcounter{footnote}{10}

\begin{abstract}
We present the first high angular resolution observations in the near-infrared $H$-band ($1.6\ \mu$m) of the 
Luminous Blue Variable star P~Cygni.  We obtained six-telescope interferometric observations 
with the CHARA Array and the MIRC beam combiner. These show that the spatial 
flux distribution is larger than expected for the stellar photosphere.  A two 
component model for the star (uniform disk) plus a halo (two-dimensional Gaussian) yields 
an excellent fit of the observations, and we suggest that the halo corresponds to 
flux emitted from the base of the stellar wind.  This wind component contributes 
about $45\%$ of the $H$-band flux and has an angular FWHM = 0.96 mas, compared to the predicted stellar diameter of 0.41 mas. 
We show several images reconstructed from the interferometric visibilities
and closure phases, and they indicate a generally spherical geometry for the wind. 
We also obtained near-infrared spectrophotometry of P~Cygni from which we derive
the flux excess compared to a purely photospheric spectral energy distribution. 
The $H$-band flux excess matches that from the 
wind flux fraction derived from the two component fits to the interferometry. 
We find evidence of significant near-infrared flux variability over the period
from 2006 to 2010 that appears similar to the variations in the H$\alpha$ emission
flux from the wind.  Future interferometric observations may be capable of recording the spatial 
variations associated with temporal changes in the wind structure. 


\end{abstract}


\section{Introduction}

Luminous Blue Variables (LBVs or S Doradus variables) are evolved, massive stars that are characterized by large mass loss rates and variability on multiple timescales.  \citet{1994PASP..106.1025H} estimate that the typical mass-loss rate of a non-erupting LBV is of the order of $\dot{M} \approx 10^{-5} - 10^{-3} M_{\odot}$ yr$^{-1}$.  The lifetime in the LBV phase is uncertain, but is on the order of 25,000 years. One of the defining criteria of the LBVs is the detection of a large-scale eruption, when the star brightens by several magnitudes. The quiescent times between these eruptions may last centuries. In addition to such rare, giant eruptions, these stars also display lesser photometric and spectroscopic variations on other timescales, ranging from days to decades or centuries \citep{2001A&A...366..508V}. The two ``prototypical" Galactic LBVs are $\eta$ Car and P Cygni, and they probably represent different extremes of both mass and mass loss rate within the scheme of LBV evolution \citep{1999SSRv...90..493I}.

The basic properties of P Cygni (HD 193237, HR 7763, Nova Cyg 1600) were estimated by \citet{1997A&A...326.1117N} and \citet{2001ASPC..233..133N} by comparing ultraviolet, optical, and infrared spectroscopic observations with results from the non-LTE atmospheric modeling code CMFGEN \citep{1998ApJ...496..407H}.  Najarro et al.~found that P Cyg has a mass-loss rate of $3.2 \times 10^{-5} M_{\odot}$ yr$^{-1}$, a terminal wind speed of $185$ km s$^{-1}$, a distance of $1.7\pm 0.1$ kpc, and an assumed continuum forming radius of 75 $R_{\odot}$. The star has a luminosity of $(5.6-7.0) \times 10^{5} L_{\odot}$, effective temperature of $(18.1-19.2)$ kK, and a gravity $\log{g_{\rm eff}} = 1.20$.  Their models of the hydrogen and helium spectral lines in the UV to NIR wavelength range yield a fractional composition of ${n_{\rm He}}/{n_{\rm H}} = 0.3$, indicating that nuclear processed gas is present in the photosphere. The model presented in \citet{2001ASPC..233..133N} also gives estimates of the abundances of C, N, O, Mg, Al, Si, Fe, Co, and Ni. Most of these elements have near solar metallicity, but C and O are depleted ($N_{\rm C}/N_{{\rm C}_\odot}=0.3$ and $N_{\rm O}/N_{{\rm O}_\odot}=0.18$) and N is enriched relative to solar ($N_{\rm N}/N_{{\rm N}_\odot}=6.5$).

\citet{2011AJ....141..120R} examined the optical variability ($V$-band photometry and H$\alpha$ spectroscopy) of P Cyg following the earlier study of \citet{2001A&A...376..898M}. Both studies found that the H$\alpha$ emission line increases and decreases in concert with the $V$-band flux. The variability in the $V$-band is due to the portion of the optical flux that originates in the stellar wind, but some variability may also be due to the pulsation properties of the star (e.g., \citealt{2001ASPC..233...31P}). Furthermore, Richardson et al.~discovered that there were ``discrete absorption components" (DACs) in the P Cygni absorption trough of the H$\alpha$ line profile that moved blueward with time, indicative of accelerating matter in our line of sight. These were long-lived features (2--3 yrs), with a possible recurrence time of $\sim4.7$ yrs.  This absorption strength increases when $V$ is brighter and H$\alpha$ is stronger. 
The absorption variations (formed in the gas column projected against the star), matched those of the continuum and line emission (formed in a large volume surrounding the star) and that was interpreted as evidence that the structure of the wind in our direction is comparable to that in other directions, i.e., that the wind is spherical in geometry. 

High angular resolution techniques can aid our understanding of the stellar winds through measurements of the geometry of the outflows.  Stellar winds around hot, massive stars are a source of infrared and radio flux from free-free and bound-free processes. Because the optical depth of the wind increases with wavelength, the observed diameter of the star will appear larger at longer wavelengths (e.g., \citealt{1999isw..book.....L}).  The advent of long-baseline optical and near-infrared interferometry provides a method of directly measuring the extent of emitting regions and determining the amount of flux excess at longer wavelengths. Such studies can be compared to models of stellar atmospheres and winds, such as those of Hillier \& Miller~(1998; CMFGEN). For example, \citet{2007A&A...464...87W} found evidence of an asymmetric wind from the enigmatic LBV $\eta$ Car using VLTI/AMBER interferometry.

Three key studies have used high angular resolution techniques to examine the H$\alpha$ emission of P Cygni. \citet{1997A&A...323..183V} used the Grand Interf{\'e}rom{\`e}tre {\`a} 2
T{\'e}lescopes to resolve an H$\alpha$ emitting region of angular diameter $5.5 \pm 0.5$ mas, which corresponds to 14 $R_{*}$ for the stellar diameter and distance derived by \citet{1997A&A...326.1117N}.   A similar result was recently obtained by \citet{2010AJ....139.2269B} based upon observations made with the Navy Precision Optical Interferometer.  They compared the angular extent of the H$\alpha$ emission region to that of the nearby continuum and then derived an H$\alpha$ emitting region that was 3--7 mas in diameter.  The H$\alpha$ emission from the outer region of the wind was explored using adaptive optics techniques by \citet{2000A&AS..144..523C}, who made a reconstructed image that shows a faint and clumpy halo with a radius of $\sim 200$ mas.  The wind structure on the largest scales was examined through radio observations by \citet{1997MNRAS.288L...7S, 1998MNRAS.296..669S}.  Their studies resolved  circumstellar structures that extend to an angular size of $\sim 1$\arcmin.  The structures show a clumpy and/or filamentary appearance.  Skinner et al.~developed a spherically symmetric wind model for the resolved radio nebula, and while they were able to obtain a good fit to the spectral energy distribution, their model did not explain the structure present in the observations. 

Here we investigate the wind of P Cygni through a comparison of H$\alpha$ spectroscopy, near infrared $H$- and $K$-band spectrophotometry, and interferometric $H$-band measurements from the CHARA Array using the MIRC beam combiner.  We outline our spectroscopic measurements in Section 2, and in Section 3 we discuss our interferometric measurements and an image reconstruction based upon these data.  We discuss these results and offer conclusions from our study in Section 4.

\section{Spectroscopy and Photometry}

We collected high resolution H$\alpha$ spectroscopy and near-infrared spectrophotometry within 2--3 weeks of our interferometric observations. The H$\alpha$ spectra were obtained from two observatories.  The first set of two spectra was obtained at the University of Toledo's Ritter Observatory with the 1~m telescope and \'echelle spectrograph \citep{1997PASP..109..676M}. The detector was a Spectral Instruments 600 Series camera, with a front-illuminated Imager Labs IL-C2004 4100$\times$4096 pixel sensor (15$\times$15 $\mu$m pixels). Wavelength calibration was accomplished with a Th-Ar discharge lamp. These high resolving power ($R = $26,000) spectra were reduced using standard techniques with IRAF\footnote{IRAF is distributed by the National Optical Astronomy Observatory, which is operated by the Association of Universities for Research in Astronomy, Inc., under cooperative agreement with the National Science Foundation.} with bias frames and flat fields obtained on the same night.  We reduced three orders of \'echelle data, which span 6285--6443 \AA, 6470--6633 \AA, and 6666--6834 \AA. The resulting S/N is roughly 100 in the continuum near H$\alpha$.

A second set of five H$\alpha$ spectra was obtained at Georgia State University's Hard Labor Creek Observatory (HLCO). The data were collected with a 0.5 m RC Optical Systems telescope\footnote{http://www.rcopticalsystems.com/telescopes/20truss.html} and an LHIRES III spectrograph\footnote{http://www.shelyak.com/rubrique.php?id\_rubrique=6\&lang=2}. These spectra were recorded on a thermo-electrically cooled SBIG-ST8XME CCD. The dispersion was accomplished with gratings of either 2400 grooves mm$^{-1}$ ($R\sim$ 18,000) or 600 grooves mm$^{-1}$ ($R \sim$ 4,500).  These data were reduced with standard techniques in IRAF utilizing bias, dark, and flat fields. Wavelength calibration was accomplished with a built-in Ne discharge lamp in the spectrograph.  

A log of the H$\alpha$ measurements is listed in Table 1, along with contemporaneous photoelectric $V$-band measurements from the American Association for Variable Star Observers (AAVSO). The equivalent widths of the H$\alpha$ profiles were measured by integration between 6510 and 6617 \AA\ for all spectra, which is the same range used by \citet{2001A&A...376..898M}.  Changes in the continuum flux can cause apparent equivalent width variations (based upon the changing ratio of emission to continuum flux), so we corrected for the variable continuum in the same manner as done by \citet{2001A&A...376..898M} and \citet{2011AJ....141..120R} using the $V$-band estimates in Table~1, i.e., $$W_{\lambda}({\rm corr})= W_\lambda({\rm net})10^{-0.4(V(t) - 4.8))}.$$ 

We obtained a single epoch of near-infrared spectrophotometry (Table 2) in the $K$- and $L$- bands in 2006 with the NASA Infrared Telescope Facility and SpeX cross-dispersed spectrograph \citep{2003PASP..115..362R}.  This observation (reported earlier by \citealt{2010PASP..122..379T}) was collected with a 3\arcsec\ wide slit.  We also obtained five epochs of near-infrared spectrophotometry in the $H$- and $K$-bands using the Mimir instrument on Lowell Observatory's Perkins telescope \citep{2007PASP..119.1385C} between 2008 and 2011. These observations were obtained with a 10\arcsec\ wide slit. These spectra were made by combining at least 10 individual spectra obtained by dithering along the slit. 

Reductions of the SpeX spectrum were carried out with {\tt SpeXtool} package (\citealt{2003PASP..115..389V}; \citealt{2004PASP..116..362C}). Reductions of the Mimir data were accomplished using custom software\footnote{Available for download at http://people.bu.edu/clemens/mimir/software.html.} that used bias, dark, and flat field frames. We corrected for the Mimir detector's non-linearity through a series of flat fields obtained on the same observing run to establish the response of each pixel to increasing exposure time \citep{2007PASP..119.1385C}. The Mimir and SpeX spectra were transformed to an absolute flux scale (and corrected for telluric absorption) using the {\it xtellcor} package \citep{2003PASP..115..389V}. This method uses flux calibrator stars of spectral type A0 V (in this case HD 192538) that are transformed to flux through reference to a model Vega spectrum calculated by R. Kurucz. The transformation takes into account rotational broadening, interstellar extinction, and the $B$ and $V$ magnitudes of the calibrator (in this case, $B=6.48$, $V=6.45$). A log of the NIR spectrophotometry is given in Table 2.  Note that the spectrophotometry obtained in 2006 and 2008 was previously published by \citet{2010PASP..122..379T}. All optical and NIR spectra are available upon request.

The $H$- and $K$-band spectra are plotted in Figure~1, and their relative flux placements confirm the temporal variability found by \citet{2010PASP..122..379T}.   Figure~1 also shows the predicted photospheric SED from a Kurucz model that was normalized to the blue part of the spectrum where the wind contributes little flux \citep{2010PASP..122..379T}.  During the course of our observations, the flux excess relative to the photospheric SED varied between 0.38 and 0.64 mag in the $H$-band and between 0.63 and 0.86 mag in the $K$-band. 

In addition to the optical H$\alpha$ spectroscopy and the NIR spectrophotometry, we collected $V$-band data from the AAVSO and the recent analysis of \citet{2012JAVSO..40..894P}. 
Figure~2 shows a comparison of the variations over the last six years in the $V$-band magnitude, H$\alpha$ emission equivalent width, and IR flux excesses.  The AAVSO photoelectric photometry (PEP) measurements agree well with the $V$-band measurements of \citet{2012JAVSO..40..894P}, and they show that a local fading occurred around 2010.0 that was followed by a gradual increase in brightness.  Some of the $V$-band measurements from \citet{2012JAVSO..40..894P} were also reported to the AAVSO, so we removed the duplicate points from the AAVSO data set shown in Figure 2. The H$\alpha$ equivalent widths from Balan et al.~(2010), \citet{2011AJ....141..120R}, and \citet{2012JAVSO..40..894P} are shown together with our new measurements in the middle panel (all corrected for the changing continuum). These show many of the same trends seen in the $V$-band data, and we show a long-term running average of the measurements in the top two panels to illustrate this. This was calculated using a Gaussian weighting scheme parameterized by a FWHM of 300 days. The lower panel shows the $H$- and $K$- band flux excesses in the same format, and these quantities show evidence that they are correlated with the $V$-band magnitudes and H$\alpha$ emission strengths (in particular showing the same fading near 2010.0).   These trends suggest that all four of these quantities vary in concert according to changes in the mass loss rate, as suggested in the analysis of \citet{2011AJ....141..120R}. With more NIR data, future analyses may show a direct correlation between the NIR and visual magnitudes. 

\section{Interferometry}

\subsection{Observations}

We obtained interferometric observations of P Cyg on three nights, two in 2010, and one in 2011, using the MIRC beam combiner \citep{2004SPIE.5491.1370M, 2006SPIE.6268E..55M} at the CHARA Array \citep{2005ApJ...628..453T}.  We observed with MIRC using the low-resolution prism ($R\sim 42$), which disperses the light across eight spectral channels in the $H$-band (1.50$-$1.75 $\mu$m; $\Delta\lambda \sim 0.034$ $\mu$m for each spectral channel).  Table 3 presents a log of the observing dates, telescope configurations, range of baseline lengths, and observed calibrators.  On 2010 Aug 21 and 23, we combined the light from three and four telescopes, respectively.  On 2011 Sept 3, we combined the light from all six telescopes simultaneously.  All observations made use of the photometric channels installed on MIRC, which measure directly the contribution of light from each telescope and improve the visibility and closure phase calibration \citep{2010SPIE.7734E..91C}.  

To measure the instrument response, we observed calibrator stars with angular diameters smaller than 0.7 mas.  On each night,  we observed $\sigma$ Cyg as the primary calibrator; we also used calibrators observed a couple of hours following the P Cyg observations to monitor the stability of the visibility calibration. The adopted limb-darkened diameters ($\theta_{LD}$) of the calibrators are listed in Table 4.  The data were reduced using the standard MIRC reduction pipeline \citep{2007Sci...317..342M}.  The visibilities and closure phases were averaged over the 2$-$3 min observing blocks.  Based on an overall assessment of the data quality obtained with MIRC using the photometric channels, we applied minimum baseline uncertainties of 5\% to the squared visibilities and $0\fdg3$  to the closure phases.  The calibrated OIFITS data files \citep{2005PASP..117.1255P} are available on request.

Figure 3 shows the $(u, v)$ coverage on the sky sampled by the CHARA Array during the three nights of the P Cyg observations obtained with MIRC.  Figure 4 shows a plot of the squared visibilities measured with MIRC during all three nights.  The visibilities drop steadily with increasing baseline, and so with spatial frequency, indicating that the object is mostly symmetric and resolved on the longest baselines.  Figure 5 shows the closure phases measured on each closed triangle.  There are small non-zero closure phases ($\sim$ 2$^\circ$) on some triangles,  possibly indicating the presence of a small asymmetry in the light distribution.

\subsection{Geometric Models}

Our results in Figure 2 show that the star's H$\alpha$ emission was stronger during the 2011 CHARA observations than during the 2010 CHARA observation. However, the visibilities measured by CHARA/MIRC for both sets in the $H$-band are comparable. Therefore, in addition to fitting the data from each epoch separately, we also combined the two CHARA data sets to constrain better the models.  As Figure 6 shows, we fit three types of geometrical models to the interferometric visiblities: a single uniform disk, a single circularly symmetric Gaussian, and a two-component model where the star is represented as a uniform disk and the extended wind emission is represented by a circular Gaussian centered on the star.  In the two-component model, we fixed the stellar disk diameter at $\theta_{\rm UD}$ = 0.411 mas (75 $R_\odot$ at 1.7 kpc), as found by \citet{1997A&A...326.1117N} and \citet{2001ASPC..233..133N}, and solved for the FWHM size of the Gaussian wind ($\theta_{\rm FWHM}$) and the flux ratio between the wind and the star ($f_{\rm wind}/f_{\rm star}$).  

Table 5 lists the results for the 2010 and 2011 epochs separately as well as for the combined data set. In Figure 4, the comparison of the models to the visibilities demonstrates the superiority of the two-component model. In both epochs as well as the combined dataset, the $\chi^2_\nu$ is significantly improved in the two-component model compared with a single uniform disk or a single circular Gaussian. We attempted to fit an elliptical Gaussian to the data as well, but we found that the $\chi^2$ of the fit was not improved significantly.  Moreover, the difference between the FWHM of the major and minor axes was less than 5\%, with a deviation of less than 1.2 $\sigma$, suggesting that the wind is essentially circular in the $H$-band continuum.

Figure 6 shows images of the best fit models for the uniform disk, circular Gaussian, and the two-component model for the combined data set. The best fit two-component model for the combined 2010 and 2011 data sets gives a wind FWHM size of 0.96 $\pm$ 0.02 mas ($R = 2.3 R_*$), contributing approximately 45\% of the total flux.  The parameter uncertainties quoted in Table~5 include the small spread in the results due to the uncertainties in the assumed photospheric angular diameter ($\triangle\theta_{\rm UD} / \theta_{\rm UD}\approx 3\%$).  Our geometric model assumes that the circular Gaussian representing the wind is centered on the star.  The star may in fact block flux from the wind behind the star.  To account for this effect, we calculated the total flux in the wind that is coincident with the star ($f_{\rm cent}$). We corrected the flux contributions by removing 0.5$f_{\rm cent}$ from the wind component and adding 0.5$f_{\rm cent}$ to the stellar flux (to roughly account for the foreground/background portions of the wind).  After applying this correction, we find that the wind contribution drops to 42\% of the total flux in the combined data set.

We attempted to fit for the uniform disk diameter of the star as a free parameter in the two-component model, but the results were not consistent between epochs, with values ranging from 0.39 to 0.62 mas.  The resolution of the CHARA Array on the longest baseline in the $H$-band is $\sim$ 0.52 mas, therefore, the stellar diameter is only marginally resolved and would require a more precise calibration of the interferometric visibilities to measure reliably.  The situation is further complicated by model degeneracies between the stellar diameter, size of the wind, and flux contribution of each component.  For instance, assuming a larger stellar angular diameter results in a model that contains a fainter, but more extended wind.  Because we fixed the stellar diameter over a narrower range of parameter space than is allowed for by the interferometeric data alone, we suspect that the uncertainties listed in Table~5 do not fully represent the model degeneracies.  Therefore, differences in the parameters between the two epochs should not be treated as significant.  Additionally, the sampling of the $(u, v)$ coverage may affect the parameters of the fit.  We note that the reduced $\chi^2_\nu$ of the combined fit is close to 1, indicating that the model does a good job reproducing the data from each epoch.  

\subsection{Image Reconstructions}

Figure 5 shows the closure phases measured with the MIRC beam combiner at the CHARA Array.  The small, but non-zero, closure phases show that there may be some structure in the wind.  The geometric models computed in Section 3.2 are point-symmetric and do not account for the non-zero closure phases.  We attempted to map the asymmetry by reconstructing images of P Cygni using the MArkov Chain IMager (MACIM; \citealt{2006SPIE.6268E..58I}) and BiSpectrum Maximum Entropy Method (BSMEM; \citealt{2008SPIE.7013E.121B}) software packages.  MACIM randomly moves flux within a pixel grid to reconstruct the image; the movement is regulated by a simulated annealing temperature.  BSMEM uses a gradient descent method to converge to the best image.  

Examples of different image reconstructions of P Cygni based on the MIRC data from 2010$-$2011 are shown in Figure~7.  We combined the epochs to maximize the $(u,v)$ coverage, but also reconstructed images for each epoch separately to avoid potentially blurring the motion of structures within the wind.  The top row of Figure~7 shows a comparison of images reconstructed using BSMEM and MACIM.  For the BSMEM reconstruction, we started with the initial image and a prior set to a 2.0 mas Gaussian. Note that the initial image defines the starting position of the flux distribution, while the prior image defines the probability of where the flux is likely to move during the reconstruction. For the MACIM reconstruction, we used our best fit uniform disk and Gaussian model as the initial image.  In both of these cases, we find a larger amount of flux in the central region of the image and a more compact size for the extended emission, compared to the two-component geometric models.  This could suggest that the boundary between the star and the wind is more blurred than in our simpler models.  However, this could also be the result of the software having difficulty reconstructing a sharper boundary between the ``edge'' of the star and the fainter, more diffuse emission.  The faint, extended tails in the north-south and east-west directions line up roughly with gaps in the $(u,v)$ coverage (compare the images with combined coverage for 2010$-$2011 in Fig. 3) and are most likely artifacts produced in the reconstruction process.  The generation of such artifacts could also be influenced by small, baseline-dependent calibration errors, such as the visibilities near $B/\lambda \sim (50-70) \times 10^{6}$ in Figure 4, which are systematically below the model fit.

The MACIM software allows for simultaneously fitting for a uniform disk while reconstructing the extended emission.  Including a uniform disk with a diameter of 0.411 mas that contributes 55\% of the flux, MACIM produced the images in the lower panels of Figure 7.  For both reconstructions, we used the Gaussian component of our two-component model as the initial image for the extended emission.  MACIM offers the flexibility of using different regularizers that reduce the weighting for non-physical images by balancing the tasks of lowering the $\chi^2$ statistic while optimizing the regularization statistic.  The image on the lower left of Figure 7 used the MACIM option of a ``compressed sensing regularizer," which minimizes spatial gradients in the image (e.g., \citealt{Donoho:2006:CS:2263438.2272089}; \citealt{candes}). For the image on the lower right, we used our Gaussian model as a prior to define the probability for where the emission might be located, in order to keep the emission more centrally located.  In both cases, the sharp edge between the star and the wind is retained because we included a fixed uniform disk in the reconstruction process.  These images show that the $H$-band emission from the wind of P Cygni is largely spherical and consistent with the overall size and shape that we derived from our geometric model.

In Figure 5, we overplotted the closure phases computed from the image in the lower-left panel of Figure 7.  All of the image reconstructions, for the combined data set as well as the individual epochs, provide reasonable fits to the small, non-zero closure phases, but unfortunately, we could not find a unique solution to describe the location and shape of a slight asymmetry in the wind.  The nearly point-symmetric morphology of P Cygni (with closure phases close to 0$^\circ$) makes it difficult to constrain the detailed structure of the wind.  Additionally, diffuse emission from an extended, Gaussian envelope is a difficult case for interferometric imaging, because the image reconstruction techniques can generate artifacts associated with gaps in the frequency coverage.  

\citet{1997A&A...323..183V} examined P Cygni interferometrically using the Grand
Interf{\'e}rom{\`e}tre {\`a} 2 T{\'e}lescopes (GI2T) in the H$\alpha$ and
\ion{He}{1}
$\lambda$ 6678 lines. They deduced that there was a structure in the wind
located
at a projected radial separation of
$R \approx 4 R_\star$ (0.8 mas) away from the star. The angular resolution
(FWHM) of
the CHARA Array with the MIRC beam combiner is $\approx 0.5$ mas, so we
should detect such
a structure if it is relatively bright compared to the star and wind.  The
closure
phase (in radians) is $\approx F_{\rm asymmetric} / {F_{\rm symmetric}}$
\citep{2007NewAR..51..604M},
and our results (see Fig.~5) show that the largest closure phase we
measure is $\sim
2^{\circ}$, or 0.035 radians.  If we assume the entire closure phase
quantity is due
to a single asymmetry in the wind (such as a blob or clump), then any such
blob
would contribute less than about 4\% of the symmetric stellar and wind flux.
According to our models of the wind halo (Fig.~8), the wind flux at $R =
0.8$ mas
is quite faint in the $H$-band (only about 1\% of the maximum light).
Consequently, observations like ours would only detect rather extreme
and isolated wind density enhancements (i.e., a distribution of clumps
surrounding
the star would yield a smaller net closure phase because the flux
distribution
would appear more symmetrical).

\subsection{Limits on the Presence of a Binary Companion}

\Citet{2010MNRAS.405.1924K} argued that 17th-Century eruptions of P Cygni might have resulted from an interaction with a B-type, main-sequence star in a 7-year, highly elliptical orbit. However, the historical light curve reported by \citet{1988IrAJ...18..163D}, which was used as a basis in these models, has been re-evaluated and has a more ``typical" appearance of an LBV eruption \citep{2011MNRAS.415..773S} when viewed with a sparsely sampled light curve.  While no periodic radial velocity variation was found in the recent high-resolution spectroscopic analysis over a 15-year period by \citet{2011AJ....141..120R}, the possibility of a binary is not eliminated given the high incidence of massive stars in binary systems \citep{2009AJ....137.3358M}.  A companion in a 7 yr orbit would probably have an angular semimajor axis of approximately 6 mas for the probable distance and mass of P~Cygni, and a binary with such a separation might be detectable in our interferometric observations.  

To examine that possibility, the high-precision closure phases and visibilities measured with the MIRC beam combiner at the CHARA Array were evaluated to place limits on the presence of a binary companion to P Cygni.  A binary star will produce a periodic signature in the visibilities and the closure phases, where the frequency of the variation depends on the separation of the components and the amplitude depends on their flux ratio \citep{2000plbs.conf....9B, 2007NewAR..51..604M}.  We focused our efforts on the data set from 2011 Sept 3, to avoid the motion of the hypothetical companion between 2010 and 2011.  Additionally, the 2011 data offers better $(u,v)$ coverage and more closure phase triangles for computing the detection limits.  

We investigated two possible scenarios. 
 In both cases, we fixed the diameter of the primary stellar component to be 0.41 mas (75 $R_{\odot}$) and assumed that the secondary is a point source.  In the first scenario, we explored whether the small, non-zero closure phases could be accounted for by a binary companion alone.  We used our two-component uniform disk and symmetric Gaussian model of the wind emission optimized for the 2011 epoch ($\theta_{\rm FWHM} = 0.898$ mas). We then fit for a binary system by searching through a grid of separations in RA and Dec.~over a range of $\pm$14 mas and solving for the secondary-to-primary flux ratio of the binary at each step. We also allowed the flux contributions from the primary star and its wind to vary in order to accommodate the additional flux from the hypothetical secondary.  We ran through two iterations of the grid search: on the first pass we used a step size of 0.2 mas in separation and allowed the RA and Dec.~ separations to vary to their best fit values; on the second pass we used a fixed step size of 0.01 mas to finely map the $\chi^2$ surface near the location of the absolute minimum. Using this approach, we found a best-fit binary solution where the agreement with the visibilities is similar to the symmetric two-component model and the $\chi^2$ calculated from the closure phases is reduced from 198.4 for the symmetric model, where we do not fit for the closure phases, to $\chi^2_{\rm CP} = 99.5$ when including a binary companion (based on 160 closure phase measurements).  However, the image reconstructions for this epoch give a $\chi^2$ calculated from the closure phases of only $\sim$ 54.8.  Therefore, while a binary companion could account for some of the non-zero closure phase signal, the image reconstructions that map the fine-scale structure in the wind do a better job of fitting the data. Based on the analysis of the binary fits with a two-dimensional Gaussian wind, we estimate that any possible binary companion must be more than 4.9 mag fainter than the central star in P Cygni or 5.6 mag fainter than the star+wind combined.
 
In the second scenario, we selected a MACIM image reconstruction of the wind assuming a uniform disk central star for the data from 2011 Sept 3 and investigated whether adding a binary component would improve the fit.  We added in a binary model to the image of the wind and searched through a grid of separations in RA and Dec.~over a range of $\pm$14 mas and solved for the secondary-to-primary flux ratio of the binary at each step while allowing the flux  contribution from the wind to vary.  We found a best-fit binary solution where the total $\chi^2$ calculated from the visibilities and closure phases is reduced from 154.3 ($\chi^2_{V^2} = 99.5, \chi^2_{\rm CP} = 54.8$) for the image+UD to $\chi^2$ = 145.5 ($\chi^2_{V^2} = 93.6, \chi^2_{\rm CP} = 51.9$) when including a binary companion (based on 120 visibilities and 160 closure phase measurements).  Performing an F-test on the ratio of the reduced $\chi^2_\nu$ values  (0.527/0.553 = 0.95), we find that this ratio can be exceeded by about 35\% of random observations, suggesting that the improvement in the fit  by adding the binary parameters is not significant.  Based on the analysis of the binary fits using the reconstructed image of the wind, we find a tighter restriction on the presence of a binary companion in that it must be more than 5.3 mag fainter than the central star or 6.0 mag fainter than the star+wind combined in the $H$-band.  

We can estimate the absolute magnitude of the LBV star in P~Cygni and then determine limits on the kinds of faint companions that remain undetected.  Based upon the $H$-band flux measurements given in Table~2 and the calibration of \citet{2003AJ....126.1090C}, we estimate that the apparent $H$-band magnitude was $3.28\pm 0.09$ during 2010 to 2011. 
Then, given the distance and reddening from \citet{1997A&A...326.1117N}, the absolute $H$-band magnitude at that time was $-8.14 \pm 0.15$ for the LBV and its wind, or $-7.49 \pm 0.15$ for the LBV alone (based upon the flux fraction from the results for the two-component model in Table 5).  The $\triangle H = 5.3$ mag limit from above then suggests that we would have probably detected any main sequence star brighter than $H=-2.2$ mag.  This magnitude corresponds approximately to that of a B1~V star \citep{2000asqu.book.....C}.  Thus, our results appear to rule out main-sequence companions of types O to B0V, but not later. 
In all cases, a secondary star very close (within a few $R_*$) to the primary would blend with the primary and would be undetectable. Future studies with more CHARA/MIRC data utilizing all six telescopes will produce stronger constraints on the absence of a companion. The number of known LBVs with a stellar companion is small \citep{2012ASSL..384..221V}. While this analysis does not disprove the \citet{2010MNRAS.405.1924K} conjecture, the improved light curve of the eruption shown by \citet{2011MNRAS.415..773S}, the lack of periodic radial velocity variation found by \citet{2011AJ....141..120R}, and this analysis point toward a single-star nature for P Cygni.
 
\section{Discussion and Conclusions}

Our CHARA Array observations provide the first high angular resolution look at the LBV P~Cygni in the $H$-band continuum. We find that the angular size of the wind is much larger than the 0.41 mas diameter predicted for the photosphere from spectral models and the distance of the star (\citealt{1997A&A...326.1117N}; \citealt{2001ASPC..233..133N}).  A spatial flux model consisting of a uniform disk photosphere and a circular Gaussian halo provides an excellent match to the interferometric observations.  The halo probably corresponds to flux emitted in the base of the stellar wind of P Cygni.  The FWHM of the halo light is approximately 1 mas, which is smaller than the 5.5 mas size found for the H$\alpha$ emission (formed over a larger wind volume because of its higher optical depth; \citealt{1997A&A...323..183V}).  

The two-component model also provides an estimate of the relative flux contributions of the star (uniform disk) and wind (circularly symmetric Gaussian), and the flux ratio of 1.36:1 (all epochs) from interferometry is consistent with that derived from spectrophotometry. 
The interferometry obtained in 2010 yields a ratio of $F_{\rm star} / F_{\rm wind} = 1.64 \pm 0.09$. This implies a predicted magnitude difference of $\triangle H = 2.5 \log [(F_{\rm wind} + F_{\rm star})/F_{\rm star}] = 0.52 \pm 0.05$ mag.  Our spectrophotometry obtained within a month of the 2010 interferometric data shows an excess of $\triangle H = 0.55 \pm 0.01$ mag (Table~2), in excellent agreement with the interferometric results.  Thus, the SED models used to determine the IR flux excess and the geometric models used in the interferometric analysis agree within the uncertainties.

Our interferometric results obtained with the MIRC instrument and the CHARA Array include precise closure phases (Fig.~5). These measurements reveal the possibility of a small amount of asymmetry present in the wind and may help probe the geometry of the stellar wind outflow. From image reconstructions, we found there may be some wind asymmetry (as indicated by small non-zero closure phases), but the wind is spherical within observational limits. It may be possible in future studies, with more complete coverage of the $(u,v)$ plane than we obtained (Fig.~3), to determine the exact location of any bright asymmetries in the wind.  The image reconstruction process could also be aided by the development of a regularizer that would keep the image smooth in azimuth in order to avoid the problem of diffuse, extended emission from mapping onto gaps in the $(u,v)$ coverage.

Models of LBV atmospheres and winds can accurately reproduce the emergent spectra (\citealt{1997A&A...326.1117N}; \citealt{1998ApJ...496..407H}). We utilized the CMFGEN model of P Cygni with the parameters of \citet{2001ASPC..233..133N} to compare the theoretical visibility curve with the observed visibility curve (Fig.~4). 
We computed the visibility curve for the derived wind radial light distribution in the $H$-band continuum for the adopted mass loss rate $\dot{M}=3.2\times 10^{-5} M_{\odot} {\rm yr}^{-1}$, a stellar radius of 76 $R_{\odot}$ and a distance of 1.7 kpc. The CMFGEN model predicts a radial profile with a half width half maximum of $R/R_\odot = 88$ (Fig.~8). However, the visibilities associated with the model made a poor
match with the observed values, so we rescaled the angular size until it best matched the visibilities (Fig.~4). This resulted in the CMFGEN model being scaled 17\% larger than expected. We overplot the visibilities of the scaled CMFGEN model as the dotted red line in Figure 4, and show the scaled model for comparison in Figure 6.
 The resulting agreement has a reduced $\chi_{\nu}^2$ of 1.5, only marginally worse than that of our two-component model ($\chi_{\nu}^2=1.1$). The major deviation in the scaled curve is at the largest baselines, where we sample the smallest angular scales.  We also used this scaled CMFGEN model as an initial image for image reconstruction, and we found that the resulting image was nearly identical to that which used our two-component model as the initial image.


We compare the calculated radial light distribution from the CMFGEN model to our radial distribution from the image reconstructions in Figure 8. The radial profile from our reconstructed images includes a sharp edge from the star. In reality, the sharp edge would be smoother due to effects of electron scattering and free-free emission. Therefore, in order to build a more realistic profile, we performed a Gaussian smoothing to the images.  The smoothed profiles appear similar to the predicted profile from the CMFGEN models but the scales of the inner light distribution differ. However, we caution that the angular resolution of our interferometry is insufficient to differentiate between the models at scales $\leq 0.5$ mas. The general agreement at large scales shows that the CMFGEN model
prediction about the spatial distribution of the wind flux is consistent with our observations.

The model radial intensity distribution had to be rescaled to fit the visibilities, which resulted in a size about 17\% larger than the predicted size. There are several plausible explanations for this difference.  First, P Cygni could be closer to us than the assumed distance of 1.7 kpc, so that its angular size appears larger. Secondly, there are a large number of hydrogen emission lines in the $H$-band (see Fig.~1) that form in the wind at larger radii than the continuum flux.  The emission lines are blended with the continuum flux in our low spectral resolution observations with MIRC ($R \sim 42$), but their net contribution may lead to an overestimate of the size of the spatial intensity
distribution compared to what would be observed for the continuum alone. Finally, it is possible that some adjustments to the wind model could account for the difference in the angular size of the wind flux (if the above
reasons are insufficient).  The mass-loss rate of P Cygni is known to vary (e.g., \citealt{2011AJ....141..120R}), and the observed H$\alpha$ strength was high during the second epoch of CHARA observations (Fig. 2). This might indicate that the actual mass loss rate was larger than the assumed model value. In addition, small changes related to the assumed velocity law and/or wind clumping factor could also lead to differences in the angular size prediction
comparable to the observed difference.

Our results represent the first images of the circumstellar environs close to the prototypical LBV, P Cygni. 
The wind appears to be spherically symmetric. The results of the
interferometric analysis and the spectroscopic analysis are mutually consistent, so long term temporal variations in wind emission should also be detected in spatial observations through interferometry. 
P Cygni should remain a prime target for monitoring with optical and NIR spectroscopy, photometry, and improved interferometry. 

\acknowledgments 
We acknowledge with thanks the variable star observations from the AAVSO International Database contributed by observers worldwide and used in this research. Support for Ritter Astrophysical Research Center during the time of the observations was provided by the National Science Foundation Program for Research and Education with Small Telescopes (NSF-PREST) under grant AST-0440784 (NDM). This work was also supported by the National Science Foundation under grants AST-0606861 and AST-1009080 (DRG). NDR gratefully acknowledges his current CRAQ postdoctoral fellowship. We are grateful for the insightful comments of A. F. J. Moffat that improved portions of the paper, discussions with Paco Najarro and Luc Dessart about spectroscopic modeling of P Cygni, and support of the MIRC 6 telescope beam combiner by Ettore Pedretti. Institutional support has been provided from the GSU College of Arts and Sciences and from the Research Program Enhancement fund of the Board of Regents of the University System of Georgia, administered through the GSU Office of the Vice President for Research. Operational funding for the CHARA Array is provided by the GSU College of Arts and Sciences, by the National Science Foundation through grants AST-0606958 and AST-0908253, by the W. M. Keck Foundation, and by the NASA Exoplanet Science Institute. We thank the Mount Wilson Institute for providing infrastructure support at Mount Wilson Observatory. The CHARA Array, operated by Georgia State University, was built with funding provided by the National Science Foundation, Georgia State University, the W. M. Keck Foundation, and the David and Lucile Packard Foundation. This research was conducted in part using the Mimir instrument, jointly developed at Boston University and Lowell Observatory and supported by NASA, NSF, and the W.M. Keck Foundation. JDM acknowledges Univ. of Michigan and NSF AST-0707927 for support of MIRC construction and observations. DPC acknowledges support under NSF AST-0907790 to Boston University. We gratefully acknowledge all of this support. This research has made use of the SIMBAD database operated at CDS, Strasbourg, France.

{\it Facilities:} \facility{Ritter Observatory}, \facility{AAVSO}, \facility{Perkins}, \facility{IRTF}, \facility{CHARA}, \facility{HLCO}

\bibliographystyle{apj} 

\bibliography{pcyg}


\begin{deluxetable}{ccccccc}
\tablecaption{H$\alpha$ Spectroscopy}
\tablewidth{0pt}
\rotate
\centering
\tablehead{
  \colhead{}         &
  \colhead{} &
  \colhead{Date} &
  \colhead{} &
  \colhead{$W_{\lambda}$}  &
  \colhead{}  &  
  \colhead{$W_{\lambda}$}  \\  

  \colhead{UT}  &
  \colhead{}  & 
  \colhead{(HJD--2,400,000)}         &
  \colhead{} &
  \colhead{(net)}  &  
  \colhead{$V$}         &  
  \colhead{(corr)}          \\ 

  \colhead{(YYYY-MM-DD)} &
  \colhead{Source} &
  \colhead{(d)}  &
  \colhead{$R$} &
  \colhead{(\AA )}  &
  \colhead{(mag)} &
  \colhead{(\AA )} \\
 
  \colhead{(1)}  &
  \colhead{(2)}  &
  \colhead{(3)} &
  \colhead{(4)} &
  \colhead{(5)} &
  \colhead{(6)} &
  \colhead{(7)} }    

\startdata
2008-10-22&  	Ritter	&54762 	&26,000	    &$-$87.4 &	4.72	&\phn $-$94.1 \\
2010-08-29&	Ritter	&55438	&26,000	    &$-$85.8 &	4.81	&\phn $-$85.0 \\
2011-05-10&	HLCO	&55692	&\phn 4,500 &$-$89.6 &	4.67	&    $-$101.0 \\
2011-05-24&	HLCO	&55706	&18,000	    &$-$86.7 &	4.67	&\phn $-$97.7 \\
2011-06-14&	HLCO	&55727	&\phn 4,500 &$-$86.2 &  4.62    &    $-$101.7 \\
2011-06-21&     HLCO    &55734  &\phn 4,500 &$-$84.1 &  4.58    &    $-$103.0 \\
2011-08-16&     HLCO    &55790  &\phn 4,500 &$-$89.6 &  4.66    &    $-$101.9 \\
\enddata
\end{deluxetable}

\clearpage
\begin{deluxetable}{ccccccc}
\tablecaption{NIR Spectrophotometry}
\tablewidth{0pt}
\tabletypesize{\footnotesize}
\rotate
\centering
\tablehead{
  \colhead{UT}         &
  \colhead{} &
  \colhead{Date} &
  \colhead{$\log (F_H)$}  &
  \colhead{$\triangle H$} &
  \colhead{$\log (F_K)$}  &
  \colhead{$\triangle K$} \\  

  \colhead{Date}  &
  \colhead{Observatory/}  & 
  \colhead{(HJD--2,400,000)}         &
  \colhead{(1.629 $\mu$m)}         &  
  \colhead{(1.629 $\mu$m)}         &
  \colhead{(2.179 $\mu$m)}        &
  \colhead{(2.179 $\mu$m)}         \\ 

  \colhead{(YYYY-MM-DD)} &
  \colhead{Instrument} &
  \colhead{(d)}  &
  \colhead{(ergs s$^{-1}$ cm$^{-2}$ \AA$^{-1}$)} &
  \colhead{(mag)} &
  \colhead{(ergs s$^{-1}$ cm$^{-2}$ \AA$^{-1}$)} &
  \colhead{(mag)}  \\
 
  \colhead{(1)}  &
  \colhead{(2)}  &
  \colhead{(3)} &
  \colhead{(4)} &
  \colhead{(5)} &
  \colhead{(6)} &
  \colhead{(7)}    }

\startdata
2006-09-16	&	IRTF/SpeX	&	53994	& \nodata	&	\nodata	&	$-11.62$	&	0.72	 \\
2008-10-20	&	Lowell/Mimir	&	54759	& $-11.21$	&	0.64 	&	$-11.57$	&	0.86	 \\
2009-07-13	&	Lowell/Mimir	&	55025	& $-11.27$	&	0.48	&	$-11.62$	&	0.73	 \\
2009-11-06	&	Lowell/Mimir	&	55141	& $-11.31$	&	0.38	&	$-11.66$	&	0.63	 \\
2010-07-03	&	Lowell/Mimir	&	55380	& $-11.25$	& 	0.55	&	$-11.59$	&	0.81	 \\
2010-11-27	&	Lowell/Mimir	&	55527	& $-11.25$	&	0.55	&	$-11.60$	&	0.80	 \\
\nodata		&	MODEL		&\nodata		& $-11.46$	&	0.00	&	$-11.91$	&	0.00	 \\
\enddata
\end{deluxetable}

\clearpage

\begin{deluxetable}{llccl}
\rotate
\centering
\tablewidth{0pt}
\tablecaption{CHARA/MIRC Observations}
\tablehead{
\colhead{UT Date} & 
\colhead{Telescope} & 
\colhead{Min. Baseline} & 
\colhead{Max. Baseline} & 
\colhead{Calibrator} \\
\colhead{(YYYY-MM-DD)} & 
\colhead{Designations} & 
\colhead{(m)} & 
\colhead{(m)} & 
\colhead{Names} \\

}
\startdata
2010-08-20 &  S1-W1-W2          & 108    & 279 & $\sigma$ Cyg, $\zeta$ Cas          \\ 
2010-08-23 &  S2-E2-W1-W2       & 108   & 251 & $\sigma$ Cyg, 7 And, $\zeta$ Cas   \\ 
2011-09-03 &  S1-S2-E1-E2-W1-W2 &\phn 34 & 331 & $\sigma$ Cyg, 7 And, $\theta$ Cas  \\ 
\enddata
\label{tab.obslog}
\end{deluxetable}

\clearpage
\begin{deluxetable}{llll}
\tablewidth{0pt}
\tablecaption{Adopted Angular Diameters for the Calibrators}
\tablehead{
\colhead{Name} & \colhead{HD} & \colhead{$\theta_{LD}$} & \colhead{Reference} \\
\colhead{    } & \colhead{  } & \colhead{(mas)}         & \colhead{         }}
\startdata
$\sigma$ Cyg & 202850 &  0.574  $\pm$ 0.017  &  1            \\ 
7 And        & 219080 &  0.659  $\pm$ 0.017  &  2, 3, 4, 5   \\ 
$\zeta$ Cas  & 3360   &  0.307  $\pm$ 0.021  &  6            \\ 
$\theta$ Cas & 6961   &  0.608  $\pm$ 0.019  &  7            \\ 
\enddata
\tablerefs{(1) \citet{2010AJ....140.1838S}; (2) \citet{2011ApJ...732...68C}; (3) \citet{1978MNRAS.183..285B}; 
 (4) \citet{2006A&A...456..789B}; (5) \citet{2008A&A...491..855K}; (6) \citet{2007A&A...464...87W};
 (7) \citet{2007ApJ...654..527G}.}
\label{tab.cal}
\end{deluxetable}

\clearpage

\begin{deluxetable}{llll}
\tablewidth{0pt}
\tablecaption{Model Fit Results}
\tablehead{\colhead{Parameter} & \colhead{UT 2010 Aug 20+23} & \colhead{UT 2011 Sep 3\phantom{20+2}} & \colhead{Combined Fit}}
\startdata
\multicolumn{4}{c}{Uniform Disk Model} \\ 
$\theta_{\rm UD}$ (mas)    & 0.902 $\pm$ 0.007   &  0.838 $\pm$ 0.004  &  0.858 $\pm$ 0.005 \\
$\chi^2_\nu$               & 6.21               &   9.28              &  9.04              \\
 \noalign{\vskip .8ex}%
 \hline
 \noalign{\vskip .8ex}%
\multicolumn{4}{c}{Circularly Symmetric Gaussian Model} \\
$\theta_{\rm FWHM}$ (mas)  &  0.558 $\pm$ 0.004  & 0.524 $\pm$ 0.003 &  0.536 $\pm$ 0.003 \\
$\chi^2_\nu$              &  4.57               &  5.83             &  5.99              \\
 \noalign{\vskip .8ex}%
 \hline
 \noalign{\vskip .8ex}%
\multicolumn{4}{c}{Two-Component Model: Uniform Disk and Circular Gaussian} \\
$f_{\rm wind}$             &  0.396 $\pm$ 0.014  & 0.475 $\pm$ 0.020  &  0.450 $\pm$ 0.018  \\
$f_{\rm corr,~wind}$ \tablenotemark{a}            &  0.379 &    0.443 &    0.423 \\
$\theta_{\rm FWHM}$ (mas)  &  1.121 $\pm$ 0.027  & 0.898 $\pm$ 0.022  &  0.964 $\pm$ 0.022  \\
$f_{\rm star}$             &  0.604 $\pm$ 0.014  & 0.525 $\pm$ 0.020  &  0.550 $\pm$ 0.018  \\
$f_{\rm corr,~star}$ \tablenotemark{a}            & 0.621   &  0.557  &   0.577 \\
$\theta_{\rm UD}$   (mas)  &  0.41 (fixed)       &  0.41 (fixed)       &   0.41 (fixed)       \\
$\chi^2_\nu$              &  0.69                &  1.13               &  1.13                \\
 \hline
 \noalign{\vskip .8ex}%
Num. Vis.                 & 120                   &  120               &  240           \\
\enddata
\tablenotetext{a}{These values correspond to the corrected flux where flux from a circular Gaussian wind behind the star would be blocked from our line of sight, so half of the flux in the Gaussian coincident with the star is subtracted from the wind and added to the star.}
\label{tab.models}
\end{deluxetable}


\clearpage
\begin{figure}
\scalebox{1.0}{\includegraphics[angle=90,width=6in]{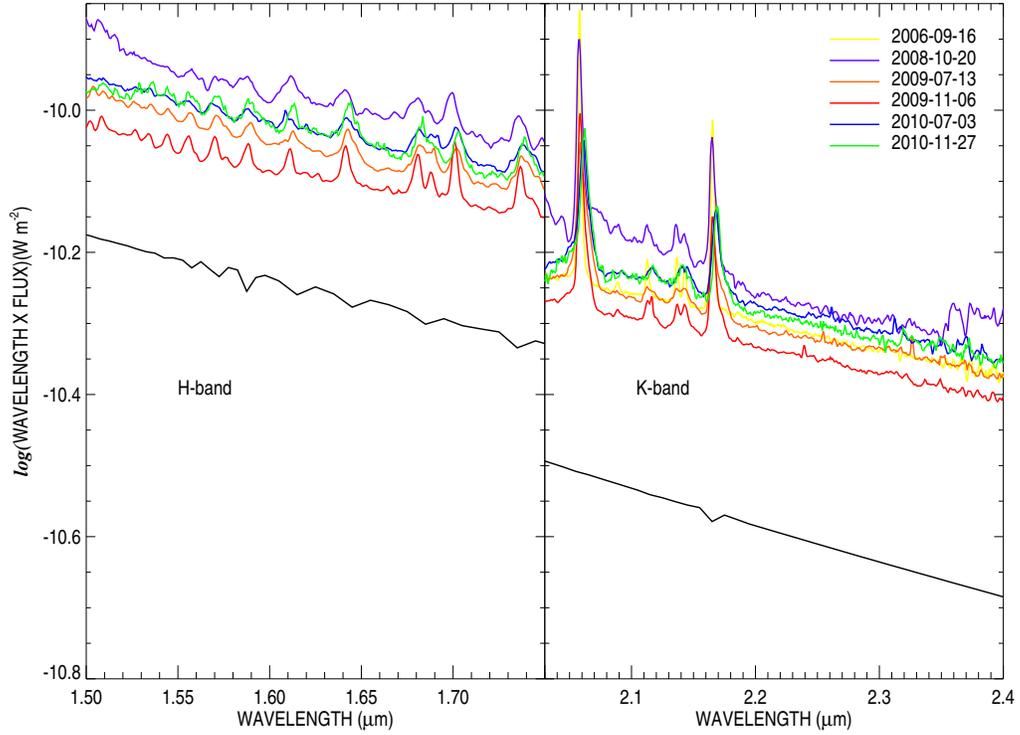}}
\caption{$H$- (left) and $K$- band (right) spectrophotometry. Dates are given in the legend and match the observations listed in Table 2. The (black) spectrum of low flux is the Kurucz model used to determine the IR excess. A color version is available in the online edition. }
\end{figure}

\clearpage
\begin{figure}
\scalebox{1.0}{\includegraphics[angle=0,width=5in]{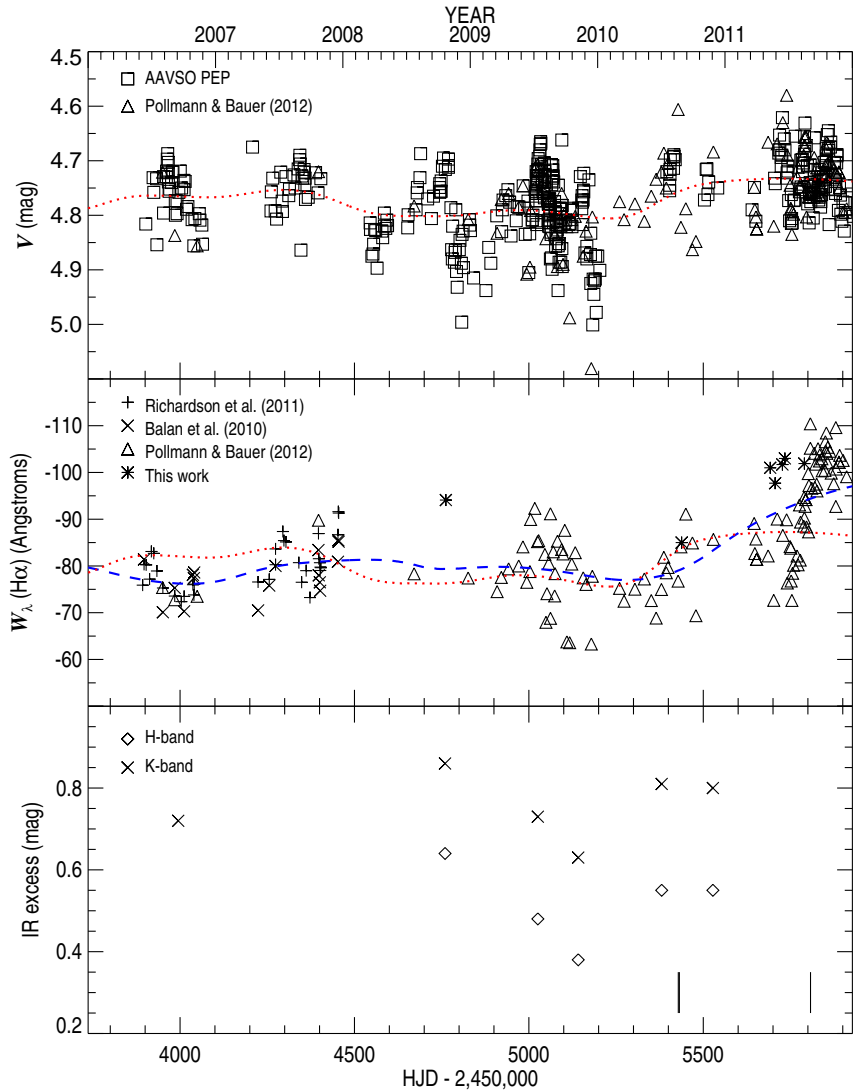}}
\caption{$V$-band photometry from the AAVSO and \citet{2012JAVSO..40..894P} is shown in the top panel (note that many of the measurements of Pollmann \& Bauer were also reported to the AAVSO, making the data sets have some overlap, so only those measurements from Pollmann \& Bauer are shown in such cases), 
H$\alpha$ equivalent width (corrected for a changing continuum) in the  middle panel, and 
the $HK$-band IR excesses in the bottom panel. We over plotted a running average of the $V$-band flux (red dotted line) in both the top and middle panel, as well as a similar curve (blue dashed line) for the H$\alpha$ measurements in the second panel to show the similarity of the variability of these measurements.
The epochs of CHARA/MIRC observations are marked with vertical lines in the bottom panel. A color version is available in the online edition.}
\label{var}
\end{figure}

\begin{figure}
\begin{center}
   \scalebox{0.5}{\includegraphics{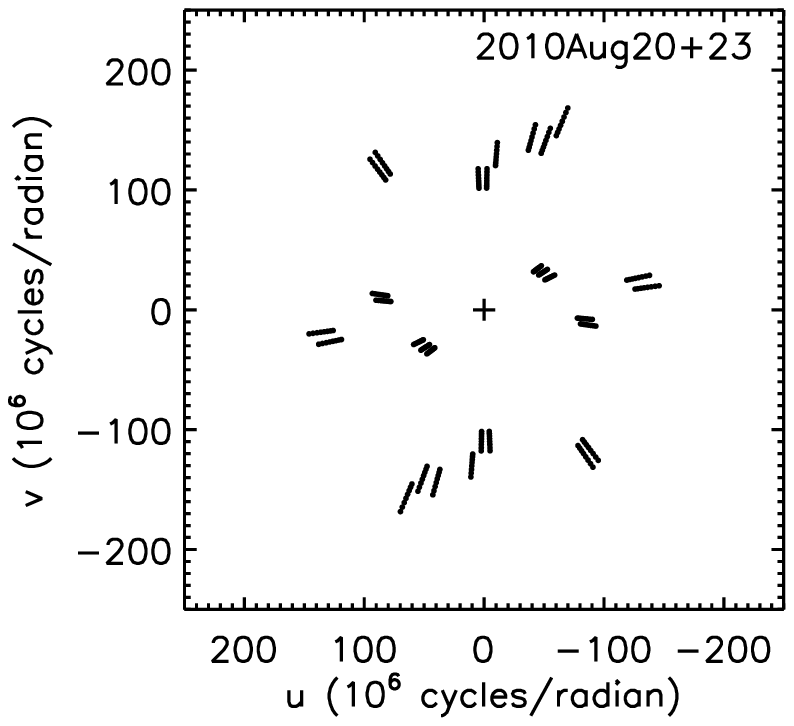}}
   \scalebox{0.5}{\includegraphics{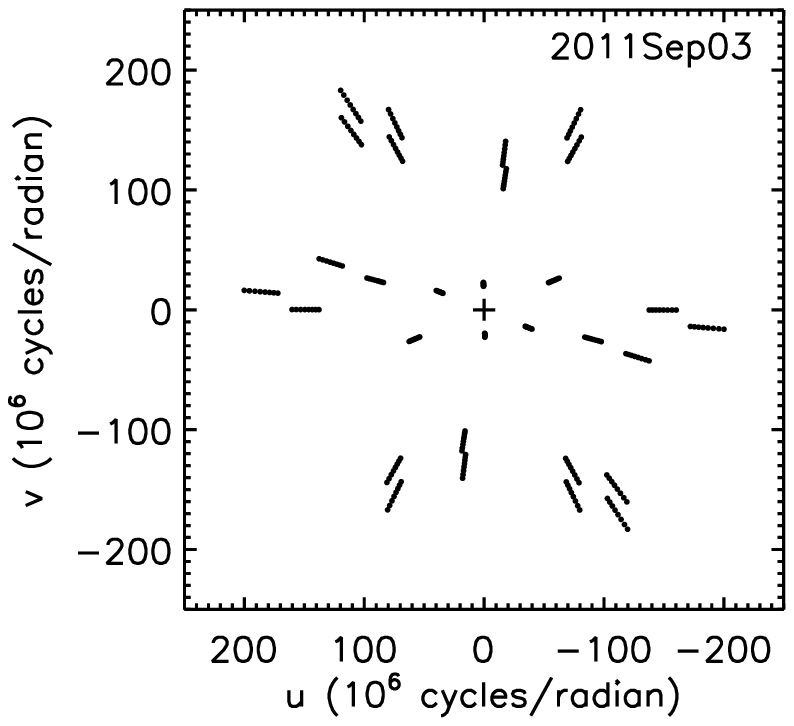}}
   \scalebox{0.5}{\includegraphics{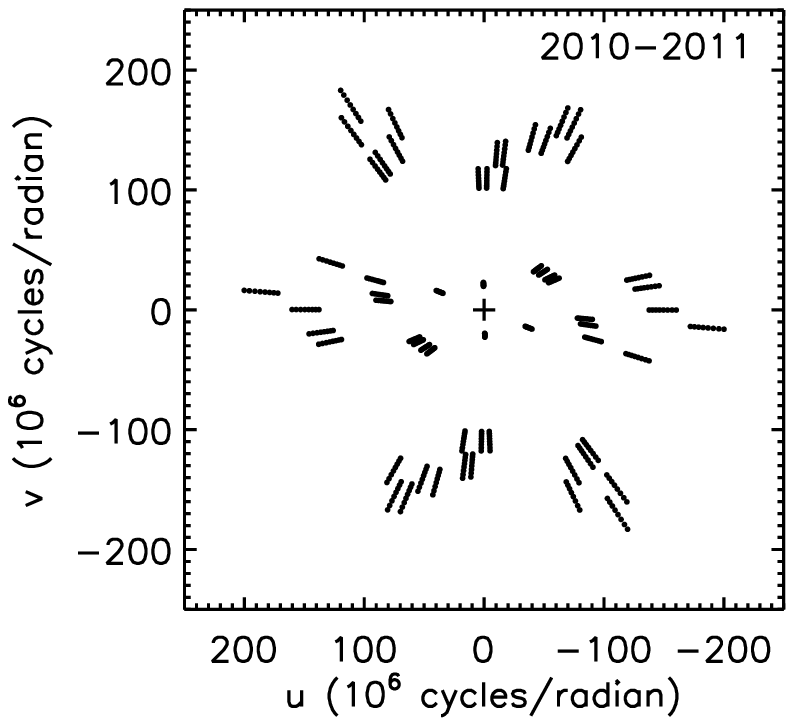}}
\end{center}
   \caption{$(u,v)$ coverage during the MIRC observations of P Cyg for the 2010 and 2011 data, 
as well as for the combined data set.}
\label{fig.uv}
\end{figure}

\clearpage

\begin{figure}
\begin{center}
   \scalebox{1.0}{\includegraphics{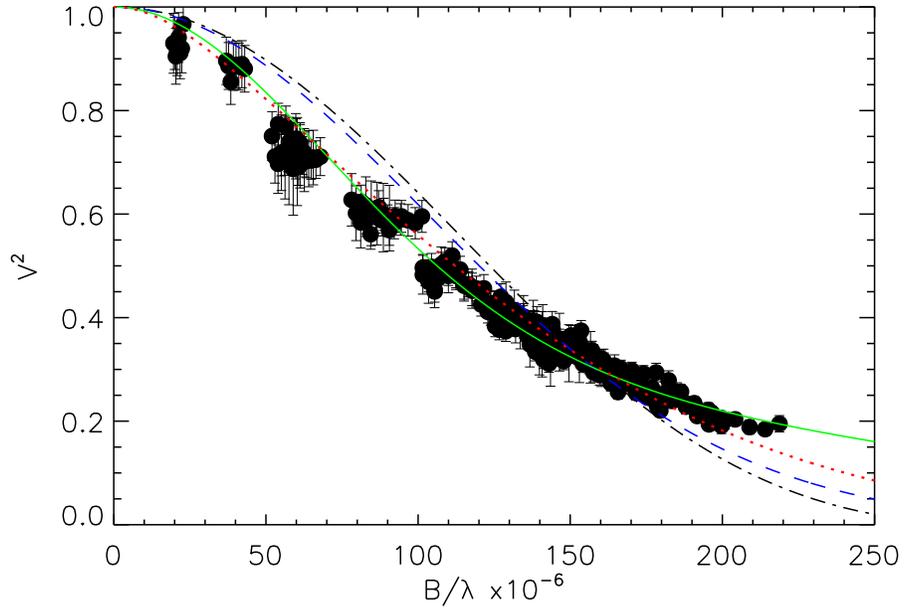}}
\end{center}
   \caption{Calibrated visibilities measured for P Cyg using MIRC on UT 2010 Aug 20+23 and 2011 Sep 3. The black dashed-dotted line is the best fit uniform disk model, the blue dashed line represents the best Gaussian model, the solid green line is our two-component model, and the dotted red line is the rescaled CMFGEN model (see discussion). A full color version is available in the online version.}
\label{fig.vis2}
\end{figure}

\clearpage

\begin{figure}
\begin{center}
\scalebox{0.5}{\includegraphics{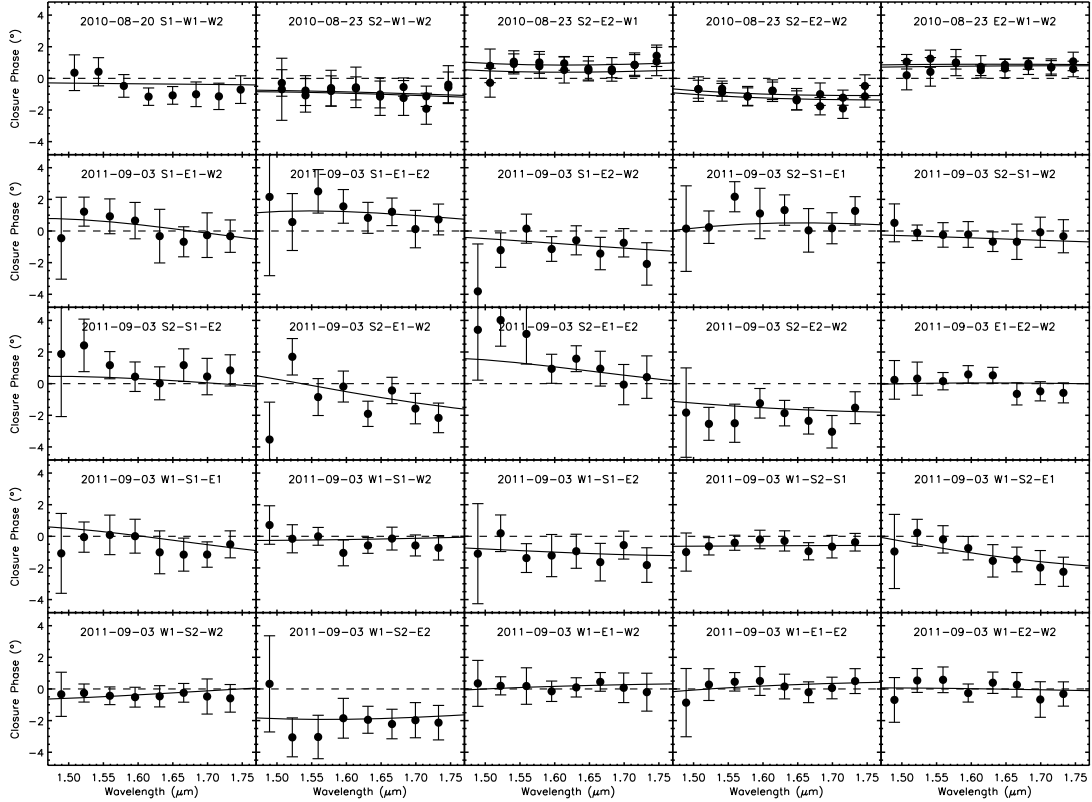}}
\caption{Closure phases measured for P Cyg using MIRC on UT 2010 Aug 20+23
and 2011 Sep 3.  The solid lines show the closure phases associated with 
the reconstructed image in the lower left panel of Figure 7. On 2010 Aug 23, we obtained two observations on P Cyg separated in time by about 30 minutes.  There is a small amount of rotation in the $(u,v)$ plane between these observations which samples the reconstructed image in a slightly different way.  The two solid lines in the plots for this date show the reconstructed closure phases for each data point.  On 2010 Aug 22 and 2011 Sep 3 we only obtained one observation of P Cyg.}
\end{center}
\end{figure}

\begin{figure}
\begin{center}
   \scalebox{0.62}{\includegraphics{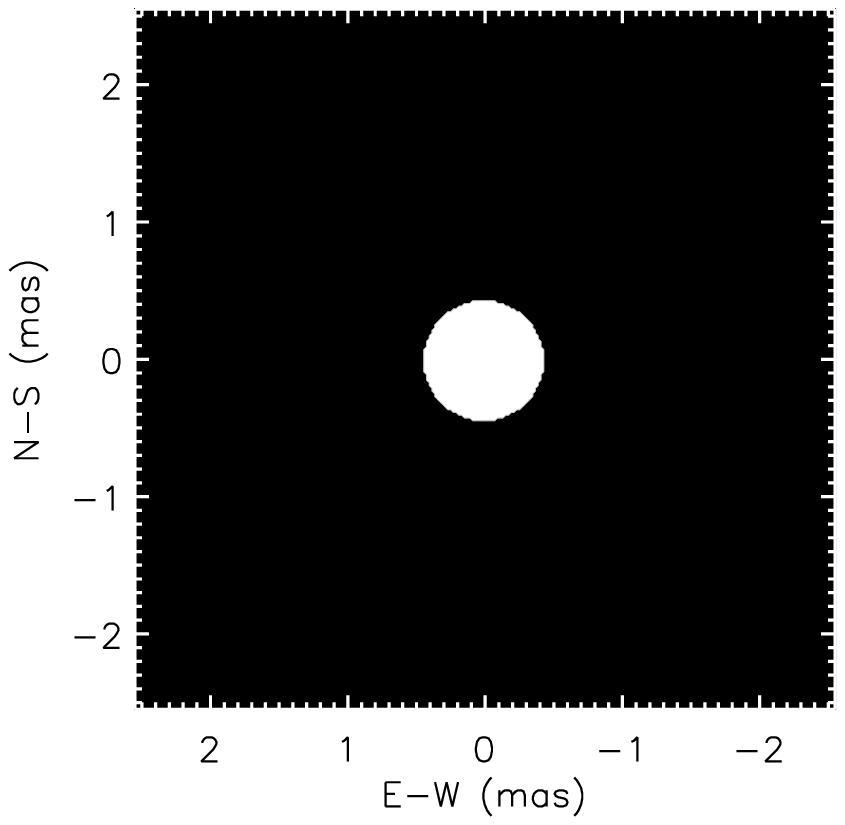}}
   \scalebox{0.62}{\includegraphics{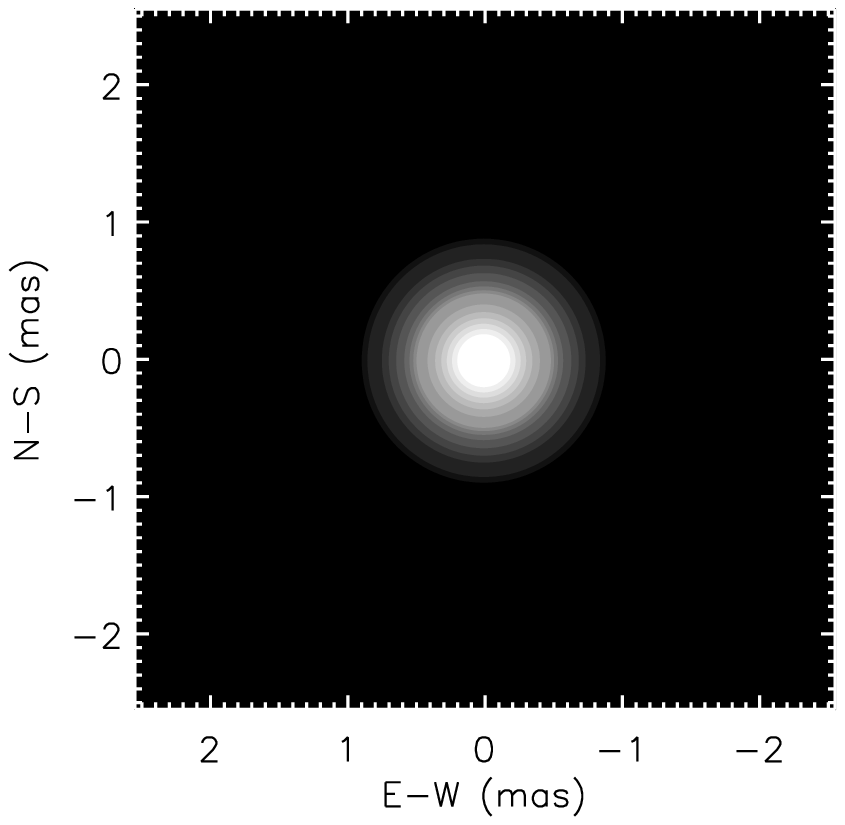}}
   \scalebox{0.62}{\includegraphics{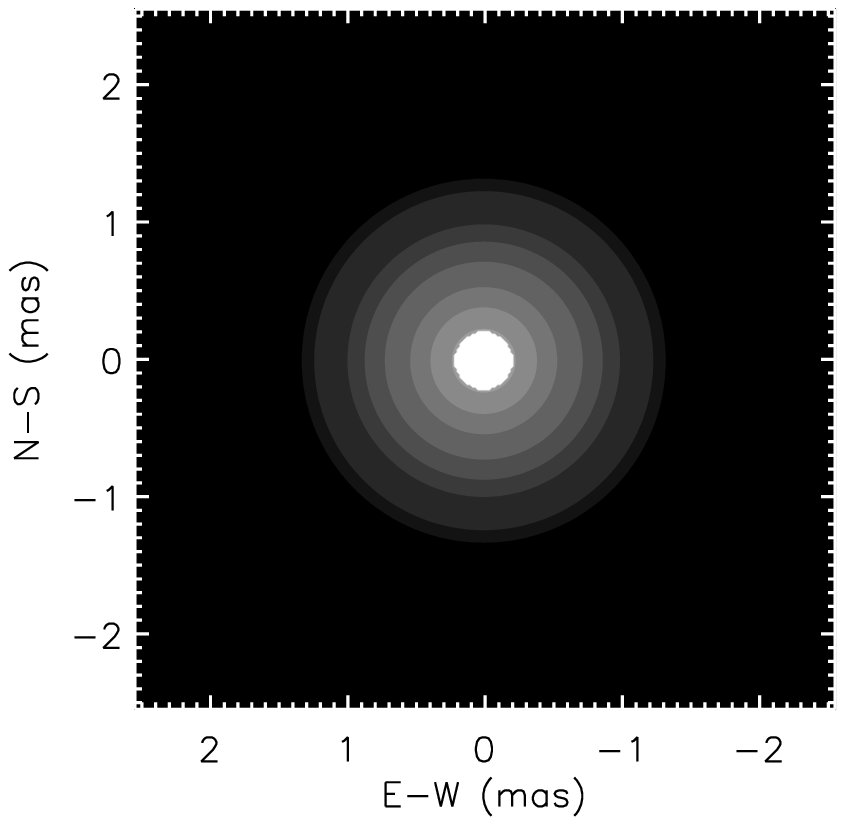}}
   \scalebox{0.62}{\includegraphics{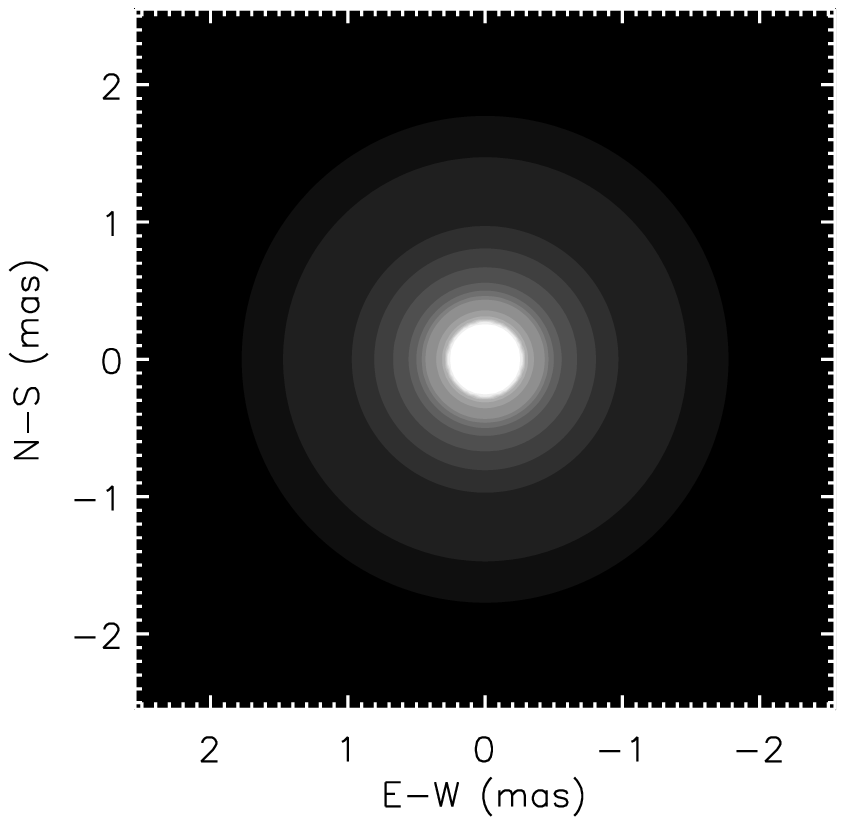}}
\end{center}
   \caption{Model flux distributions of P Cyg fit to the interferometric visibilities in the combined data from 2010-2011 : 
a uniform disk (upper left), a circularly symmetric Gaussian (upper right), and a two-component model (bottom left).  In the bottom right we show the flux distribution of the CMFGEN model scaled to optimize the fit to the interferometric visibilities (the visibility curve shown in Fig.~4).  Contour intervals are drawn at 0.01\%, 0.05\%, 0.1\%, 0.5\%, 1\%, 2\%, 4\%, 6\%, 8\%, 10\%, 20\%, 30\%, 40\%, and 50\% of the peak flux in each panel (contours extend up to 60\% and 70\% of the peak flux for the circular Gaussian and 60\%, 70\%, and 80\% of the peak flux for the CMFGEN model).}
\label{fig.models}
\end{figure}

\begin{figure}
\begin{center}
   \scalebox{0.62}{\includegraphics{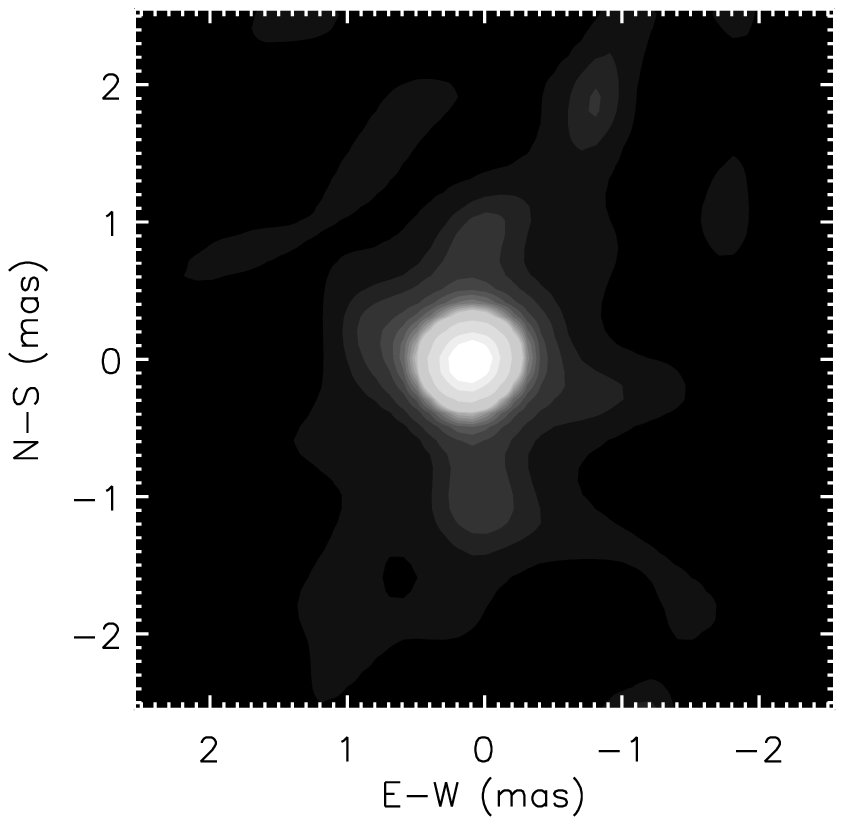}}
   \scalebox{0.62}{\includegraphics{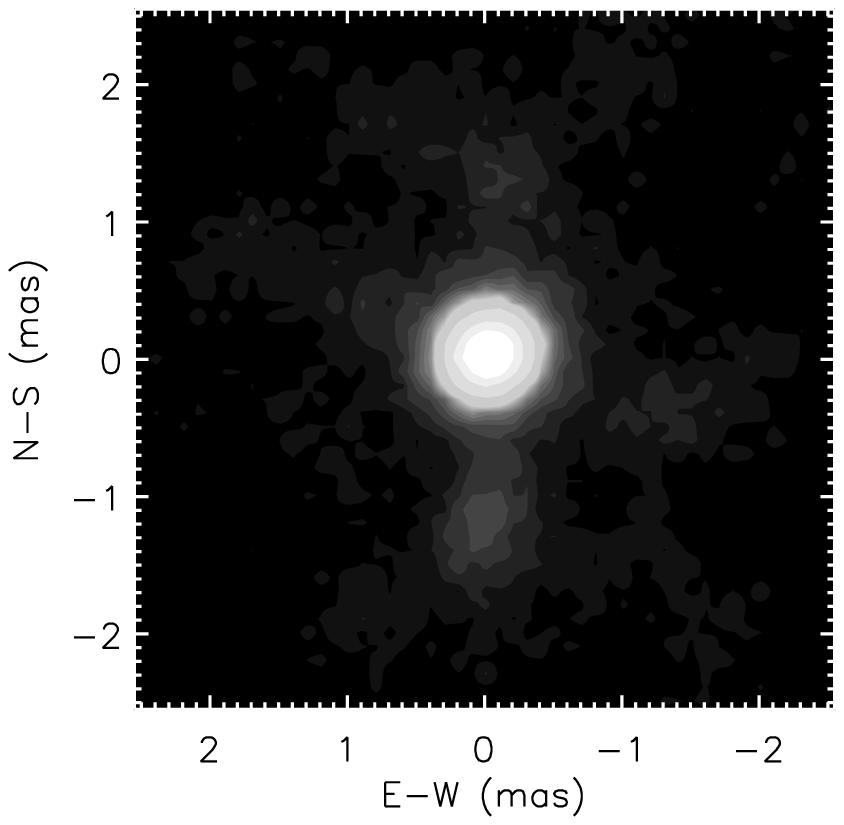}} \\
   \scalebox{0.62}{\includegraphics{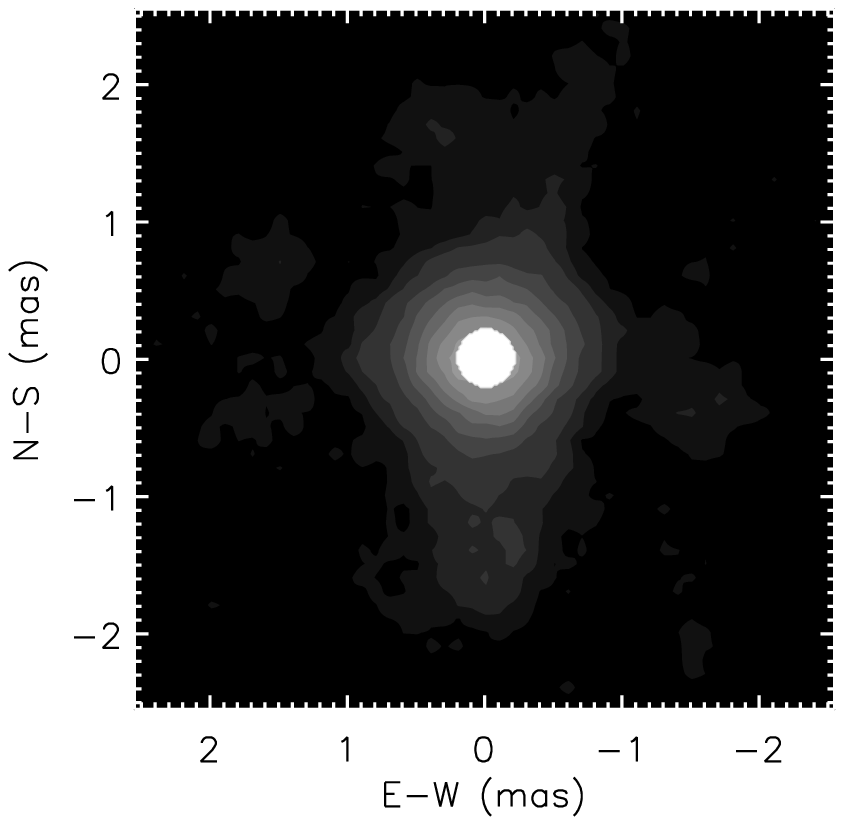}}
   \scalebox{0.62}{\includegraphics{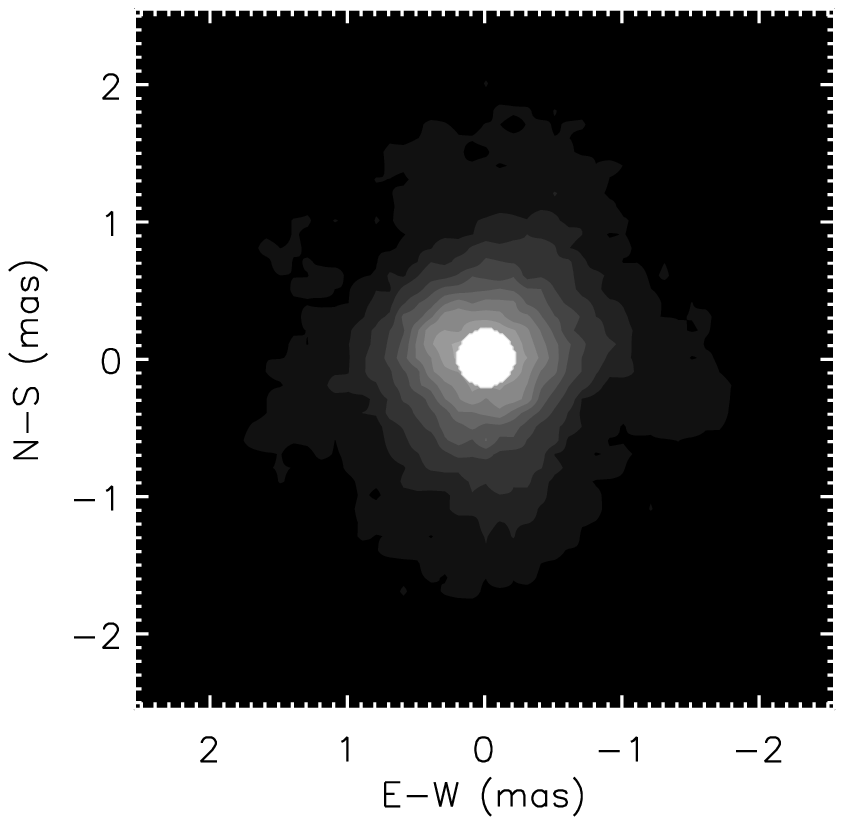}}
\end{center}
   \caption{Image reconstructions of P Cygni based on our MIRC data from 2010 and 2011 (240 visibility measurements and 232 closure phase measurements).  
Top left: Image reconstruction from BSMEM using a 2.0 mas Gaussian as the initial image and prior ($\chi^2_{V^2}$ = 331.4 for the visibilities, $\chi^2_{\rm CP}$ = 70.4 for the closure phases, where the $\chi^2$ is calculated as the the number of observations minus
the degree of the fit). 
Top right:  Image reconstruction from MACIM using our two-component geometric model as the initial image ($\chi^2_{V^2}$ = 267.1, $\chi^2_{\rm CP}$ = 82.1).  
Bottom left: MACIM image reconstruction of the extended emission while fitting for a uniform disk of 0.41 mas that contributes 55\% of the light ($\chi^2_{V^2}$ = 218.7, $\chi^2_{\rm CP}$ = 86.1).  This reconstruction used the Gaussian component of our two component model as the initial image and used a regularizer to minimize spatial gradients in the reconstructed image.
Bottom right:  MACIM image reconstruction of the extended emission made assuming a stellar flux component with a UD of 0.41 mas that contributes 55\% of the light ($\chi^2_{V^2}$ = 194.9, $\chi^2_{\rm CP}$ = 103.1).  This reconstruction used the Gaussian component from our two-component model as the initial image and as a prior to define the probability for how the flux moves during the reconstruction process.
Contour intervals are drawn at 0.05\%, 0.1\%, 0.5\%, 1\%, 2\%, 3\%, 4\%, 5\%, 6\%, 7\%, 8\%, 9\%, 10\%, 20\%, 40\%, 60\% of the peak flux in each panel.  In all cases, the $\chi^2$ calculated from the closure phases is smaller than the number of measurements.  However, we suspect that this is the result of the closure phases being so close to 0 and that small movements in the flux during the reconstruction process can reproduce the signal in many different ways, allowing the software to find a very precise, but not necessarily reliable, solution.}
\label{fig.images}
\end{figure}

\begin{figure}
\begin{center}
 \includegraphics[angle=90,width=6in]{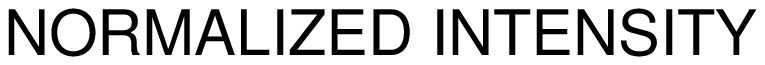}
\end{center}
   \caption{Theoretical light distribution of the star and wind in the $H$-band from CMFGEN models (red solid curve) directly from the CMFGEN model, as well as a rescaled version from a fit to the visibility curve (red dotted curve). The average radial profile of the MACIM image reconstruction (lower left panel of Fig.~7) is shown by the diamond symbols, along with Gaussian-smoothed versions of the reconstruction over 10, 20, and 30\% of the stellar radius (blue dashed lines). The purple dot-dashed curve represents the two-component model derived from the visibilities. Finally, the radial distribution of the flux from the image reconstruction with no regularizer and no uniform disk component (top right image of Fig.~7) is plotted as plus signs.  }
\label{fig.uv}
\end{figure}

\end{document}